\documentclass[12pt]{article}
\usepackage{amssymb}

\input{tcilatex}

\begin{document}

\bigskip\ 

\bigskip\ 

\begin{center}
\textbf{MAXIMAL SUPERSYMMETRY IN ELEVEN-DIMENSIONAL}

\smallskip\ 

\textbf{SUPERGRAVITY REVISITED AND CHIROTOPES}

\textbf{\ }

\smallskip\ 

J. A. Nieto \footnote{%
nieto@uas.uasnet.mx}

\smallskip

\textit{Facultad de Ciencias F\'{\i}sico-Matem\'{a}ticas, Universidad Aut%
\'{o}noma}

\textit{de Sinaloa, C.P. 80000, Culiac\'{a}n, Sinaloa, M\'{e}xico}

\bigskip\ 

\bigskip\ 

\textbf{Abstract}
\end{center}

We analyze maximal supersymmetry in eleven-dimensional supergravity from the
point of view of the oriented matroid theory. The mathematical key tools in
our discussion are the Englert solution and the chirotope concept. We argue
that chirotopes may provide other solutions not only for eleven-dimensional
supergravity but for any higher dimensional supergravity theory.

\bigskip\ 

\bigskip\ 

\bigskip\ 

\bigskip\ 


Pacs numbers: 04.60.-m, 04.30.-w, 6.20.Jr, 98.80.-k

March, 2006

\newpage \noindent \textbf{1.- Introduction}

\smallskip\ 

The main purpose of this brief note is to discuss the importance of applying
the oriented matroid theory [1] to supergravity theories. In particular, we
focus on the possibility of relating the chirotope concept [1] of oriented
matroid theory to the Freund-Rubin-Engler solution of eleven-dimensional
supergravity [2]-[4].

As it is known, the Freund-Rubin and Englert solutions compactify $d=11$
spacetime $M^{11}$ into a product of two spaces: $4$-dimensional anti-de
Sitter manifold $AdS_{4}$ and the seven sphere $S^{7}$. It turns out that
while the Freund-Rubin solution corresponds to maximally supersymmetric
solution preserving the full supersymmetry of eleven-dimensional
supergravity action, and in that sense can be considered as a "trivial"
solution [5]-[6], the Englert solution leads to spontaneous breakdown of
maximal supersymmetry and therefore can be interpreted as a "non-trivial
solution" associated with $S^{7}$-geometry.

A key object in the above solutions is a four-form field strength $F=dA$ or $%
F^{ABCD},$ with $A,B,C,D=0,...,10$. In fact, if one assumes that the only
non-vanishing components of $F^{ABCD}$ are proportional to the completely
antisymmetric symbol $\varepsilon ^{\mu \nu \alpha \beta }$, with $\mu ,\nu
,\alpha ,\beta =0,...,3,$ then the trivial solution arises from the bosonic
sector of eleven-dimensional supergravity field equations. While if in
addition one assumes non-vanishing values for $F^{ijkl}$, with $%
ijkl=4,...,10 $, one obtains the non-trivial solution. From this perspective
it becomes evident that it is important to study, deeply, the algebraic
properties of $F^{ABCD}$ and their relation with the trivial and non-trivial
solutions.

One can observe, for instance, that since in the case of maximally
supersymmetric solutions $F^{ABCD}$ is decomposable, it must be possible to
relate it to the chirotope concept via the Grassmann-Pl\"{u}cker relations
(see Ref. [7]). Here, we are interested in investigating the connection
between the object $F^{ABCD}$ associated with the Englert solution and the
chirotope concept. We argue that, in principle, our analysis can be used to
find new solutions for eleven-dimensional supergravity.

It is worth mentioning that the oriented matroid theory has been connected
with a number of topics, including $p$-branes [8], Chern-Simons theory [9],
superstrings [10], gravity [11] and two time physics [12]. In particular,
using the phirotope concept [13]-[15], which is a generalization of
chirotopes concept, in Ref. [7] a relation with super$p$-branes has been
established. These progresses have motivated a proposal of considering the
oriented matroid theory as a mathematical framework for $M$-theory [16].

In this paper, after a brief review of maximally supersymmetric solution, we
review the Englert solution putting special emphasis in the algebraic
identities of the structure constants for octonions which allow a connection
with the oriented matroid theory via the chirotope concept. Specifically, we
prove that not only in the case of the Freund-Rubin solution of
eleven-dimensional supergravity the four-form field $F$ admits an
interpretation of chirotope but also in the case of the Englert solution.
The key idea in this proof is that both sectors of $D=11$ supergravity, $4$%
-dimensional anti-de Sitter manifold $AdS_{4}$ and the seven sphere $S^{7},$
admit a realizable chirotope interpretation, although the full four-form
field $F$ may correspond to nonrealizable chirotopes. Since to each
chirotope one can associate its dual we find that our investigation may open
the possibility to find dual solutions of $D=11$ supergravity.

\bigskip\ 

\noindent \textbf{2.- Maximally Supersymmetric Solutions}

\smallskip\ 

Let $(M^{11},g,F)$ be a maximally supersymmetric solution of eleven
dimensional supergravity. In the non-degenerate case, Figueroa-O'Farrill and
Papadopoulos proved the theorem [5] that such a solution must be isometric
to either $AdS_{4}\times S^{7}$ or $AdS_{7}\times S^{4}$. Their starting
point in this result was the vanishing of the curvature $\mathcal{R}$ of the
supercovariant connection $\mathcal{D}$. In fact, demanding the vanishing of
the curvature $\mathcal{R}$ they found that $(M^{11},g,F)$ is maximally
supersymmetric solution if and only if $(M^{11},g)$ is locally symmetric
space and $F$ is parallel and decomposable, and from this results such a
theorem follows (see Ref. [5] for details).

Here, we are interested in revisiting the fact that $F$ is decomposable.
From the formula $\mathcal{R}=0$ one can essentially derive two algebraic
formulae for $F$, namely

\begin{equation}
F\wedge F=0  \tag{1}
\end{equation}%
and

\begin{equation}
_{\iota _{X}}F\wedge _{\iota _{Y}}F=0,  \tag{2}
\end{equation}%
where $\iota _{X}$ and $\iota _{Y}$ denote an inner product for the two
vectors $X$ and $Y,$ respectively. From these two formulae one then shows
that $F$ satisfies the relation

\begin{equation}
_{\iota _{Z}\iota _{Y}\iota _{X}}F\wedge F=0.  \tag{3}
\end{equation}%
Conversely, if (3) is satisfied then (1) and (2) follow. The formula (3)
means that $F$ is decomposable, that is, (3) implies that $F$ can be written
as the wedge product of four one-forms.

The way that Figueroa-O'Farrill and Papadopoulos prove that (1) and (2)
imply (3) is by first observing that contracting (1) with respect to the
three vectors $X,Y$ and $Z$ one obtains

\begin{equation}
_{\iota _{Z}\iota _{Y}\iota _{X}}F\wedge F=-_{\iota _{Y}\iota _{X}}F\wedge
_{\iota _{Z}}F.  \tag{4}
\end{equation}%
While, contracting equation (2) with a third vector field one gets

\begin{equation}
_{\iota _{Y}\iota _{X}}F\wedge _{\iota _{Z}}F=_{\iota _{Y}\iota _{Z}}F\wedge
_{\iota _{X}}F.  \tag{5}
\end{equation}%
Thus, comparing (4) and (5) one sees that whereas (5) implies that the
expression $_{\iota _{Y}\iota _{X}}F\wedge _{\iota _{Z}}F$ is symmetric in $X
$ and $Z$, (4)\ means that it is skew-symmetric. This implies that the term $%
_{\iota _{Y}\iota _{X}}F\wedge _{\iota _{Z}}F$ must vanish.

\bigskip\ 

\noindent \textbf{3.- Figueroa-O'Farrill-Papadopoulos formalism revisited}

\smallskip\ 

Let us first write the algebraic expressions (1), (2) and (3) in the
alternative way

\begin{equation}
F_{[A_{1}A_{2}A_{3}A_{4}}F_{B_{1}B_{2}B_{3}B_{4}]}=0,  \tag{6}
\end{equation}

\begin{equation}
F_{A_{1}[A_{2}A_{3}A_{4}}F_{B_{1}B_{2}B_{3}]B_{4}}=0,  \tag{7}
\end{equation}%
and%
\begin{equation}
F_{A_{1}A_{2}A_{3}[A_{4}}F_{B_{1}B_{2}B_{3}B_{4}]}=0,  \tag{8}
\end{equation}%
respectively. Here, the bracket $[,]$ means completely antisymmetric.

If we are not interested in using differential forms notation as in section
2, the contraction of (6) and (7) with respect to different vectors forces
to us to define the bracket $[,]$ in the following form

\begin{equation}
G_{[A_{1}...A_{d+1}]}\equiv G_{C_{1}...C_{d+1}}\delta
_{A_{1}...A_{d+1}}^{C_{1}...C_{d+1}}.  \tag{9}
\end{equation}%
Here, $G_{C_{1}...C_{d+1}}$ is any $d+1$-rank tensor and $\delta
_{A_{1}...A_{d+1}}^{C_{1}...C_{d+1}}$ is the generalized delta symbol. The
advantage of this notation is that several properties of the generalized
delta can be used. For instance, considering the fact that

\begin{equation}
\delta _{A_{1}...A_{d+1}}^{C_{1}...C_{d+1}}=\delta _{A_{1}}^{C_{1}}\delta
_{A_{2}...A_{d+1}}^{C_{2}...C_{d+1}}+\tsum\limits_{k=2}^{d+1}(-1)^{k}\delta
_{A_{k}}^{C_{1}}\delta _{A_{2}...\hat{A}_{k}...A_{d+1}}^{C_{2}...C_{d+1}}, 
\tag{10}
\end{equation}%
where $\hat{A}_{k}$ means omitting this index, one can easily prove that (6)
is satisfied if and only if

\begin{equation}
F_{A_{1}C_{2}C_{3}C_{4}}F_{D_{1}D_{2}D_{3}D_{4}}\delta
_{A_{2}A_{3}A_{4}B_{1}B_{2}B_{3}B_{4}}^{C_{2}C_{3}C_{4}D_{1}D_{2}D_{3}D_{4}}=0,
\tag{11}
\end{equation}%
which is equivalent to

\begin{equation}
F_{A_{1}[A_{2}A_{3}A_{4}}F_{B_{1}B_{2}B_{3}B_{4}]}=0.  \tag{12}
\end{equation}%
In the notation of section 2 this result can be written as

\begin{equation}
_{\iota _{X}}F\wedge F=0.  \tag{13}
\end{equation}

Properly, applying (10) to the expression (11) once again we get

\begin{equation}
F_{A_{1}[A_{2}A_{3}A_{4}}F_{B_{1}B_{2}B_{3}B_{4}]}=3F_{A_{1}A_{2}[A_{3}A_{4}}F_{B_{1}B_{2}B_{3}B_{4}]}+4F_{A_{1}[A_{3}A_{4}B_{1}}F_{B_{2}B_{3}B_{4}]A_{2}}.
\tag{14}
\end{equation}%
Thus, considering the fact that the second term of this expression
corresponds to formula (2) we can set

\begin{equation}
F_{A_{1}[A_{3}A_{4}B_{1}}F_{B_{2}B_{3}B_{4}]A_{2}}=0,  \tag{15}
\end{equation}%
and therefore using (12) we obtain the result

\begin{equation}
F_{A_{1}A_{2}[A_{3}A_{4}}F_{B_{1}B_{2}B_{3}B_{4}]}=0.  \tag{16}
\end{equation}

Similar technique leads to the identity

\begin{equation}
F_{A_{1}A_{2}[A_{3}A_{4}}F_{B_{1}B_{2}B_{3}B_{4}]}=2F_{A_{1}A_{2}A_{3}[A_{4}}F_{B_{1}B_{2}B_{3}B_{4}]}+4F_{A_{1}A_{2}[A_{4}B_{1}}F_{B_{2}B_{3}B_{4}]A_{3}}.
\tag{17}
\end{equation}%
Thus, using (16) we find

\begin{equation}
F_{A_{1}A_{2}A_{3}[A_{4}}F_{B_{1}B_{2}B_{3}B_{4}]}=-2F_{A_{1}A_{2}[A_{4}B_{1}}F_{B_{2}B_{3}B_{4}]A_{3}}.
\tag{18}
\end{equation}%
This expression implies that the right hand side of (18) is antisymmetric in
the indices $A_{1}$ and $A_{3}$.

On the other hand we have

\begin{equation}
\begin{array}{c}
F_{A_{1}[A_{2}A_{4}B_{1}}F_{B_{2}B_{3}B_{4}]A_{3}}=3F_{A_{1}A_{2}[A_{4}B_{1}}F_{B_{2}B_{3}B_{4}]A_{3}}-3F_{A_{1}[A_{4}B_{1}B_{2}}F_{B_{3}B_{4}]A_{2}A_{3}}
\\ 
\\ 
=3F_{A_{1}A_{2}[A_{4}B_{1}}F_{B_{2}B_{3}B_{4}]A_{3}}-3F_{A_{3}A_{2}[A_{4}B_{1}}F_{B_{2}B_{3}B_{4}]A_{1}}.%
\end{array}
\tag{19}
\end{equation}%
From (15) we see that the left hand side of (19) vanishes and therefore we
obtain

\begin{equation}
F_{A_{1}A_{2}[A_{4}B_{1}}F_{B_{2}B_{3}B_{4}]A_{3}}=F_{A_{3}A_{2}[A_{4}B_{1}}F_{B_{2}B_{3}B_{4}]A_{1}}.
\tag{20}
\end{equation}%
This means that $F_{A_{1}A_{2}[A_{4}B_{1}}F_{B_{2}B_{3}B_{4}]A_{3}}$ is
symmetric in the indices $A_{1}$ and $A_{3}$ which contradicts the
conclusion below (18). Thus we have found that the only consistent
possibility is to set

\begin{equation}
F_{A_{1}A_{2}A_{3}[A_{4}}F_{B_{1}B_{2}B_{3}B_{4}]}=0.  \tag{21}
\end{equation}%
Summarizing, we have shown that (12) and (15) imply (21). Conversely, using
once again the properties of the delta generalized $\delta
_{A_{1}...A_{d+1}}^{C_{1}...C_{d+1}}$ one can show that both $%
F_{A_{1}[A_{2}A_{3}A_{4}}F_{B_{1}B_{2}B_{3}B_{4}]}$ and $%
F_{A_{1}[A_{3}A_{4}B_{1}}F_{B_{2}B_{3}B_{4}]A_{2}}$ can be written in terms
of $F_{A_{1}A_{2}A_{3}[A_{4}}F_{B_{1}B_{2}B_{3}B_{4}]}$ and therefore (21)
implies (12) and (15). This means that the expression (21) is equivalent to
the two formulae (12) and (15). Thus, we have complete an alternative proof
of such a equivalence.

The formula (21) implies that $F_{A_{1}A_{2}A_{3}A_{4}}$ is decomposable.
This means that $F_{A_{1}A_{2}A_{3}A_{4}}$ can be written in the form

\begin{equation}
F^{A_{1}A_{2}A_{3}A_{4}}=\varepsilon
^{a_{1}a_{2}a_{3}a_{4}}v_{a_{1}}^{A_{1}}v_{a_{2}}^{A_{2}}v_{a_{3}}^{A_{3}}v_{a_{4}}^{A_{4}},
\tag{22}
\end{equation}%
where $v_{a}^{A}$ is an arbitrary $4\times d+1-$matrix. Thus, one may
conclude, as Figueroa-O'Farrill and Papadopoulos did\textbf{, } that the
maximal supersymmetry of eleven-dimensional supergravity implies that $%
F^{A_{1}A_{2}A_{3}A}$ can be written as (22). In the non-degenerate case,
spontaneous compactification allows to assume that the only nonvanishing
components of $v_{a}^{A}$ are $v_{a}^{\mu }\sim \delta _{a}^{\mu }$, with $%
\mu =0,1,2,3$ or $v_{a}^{\hat{\mu}}\sim \delta _{a}^{\hat{\mu}}$, with $\hat{%
\mu}=8,9,10,11$ leading to the two possible solutions $AdS_{4}\times S^{7}$
or $AdS_{7}\times S^{4}$, respectively (see Ref. [5] for details). In fact,
in the first case one gets from (22) that the only nonvanishing components
of $F^{A_{1}A_{2}A_{3}A_{4}}$ are $F^{\mu \nu \alpha \beta }\sim \varepsilon
^{\mu \nu \alpha \beta }$ and therefore one obtains from the
eleven-dimensional field equations

\begin{equation}
\begin{array}{c}
\frac{1}{3!}\varepsilon
_{A_{1}A_{2}A_{3}A_{4}B_{1}B_{2}B_{3}B_{4}NPQ}F^{NPQM};_{M}=\frac{1}{%
2(4!)^{2}}F_{[A_{1}A_{2}A_{3}A_{4}}F_{B_{1}B_{2}B_{3}B_{4}]}, \\ 
\\ 
R_{MN}-\frac{1}{2}g_{MN}R=\frac{1}{6}F_{MPQR}F_{N}^{PQR}-\frac{1}{48}%
g_{MN}F_{SPQR}F^{SPQR},%
\end{array}
\tag{23}
\end{equation}%
the Freund-Rubin solution $AdS_{4}\times S^{7}$. While in the second case $%
F^{\hat{\mu}\hat{\nu}\hat{\alpha}\hat{\beta}}\sim \varepsilon ^{\hat{\mu}%
\hat{\nu}\hat{\alpha}\hat{\beta}}$, the field equations (23) lead to the
solution $AdS_{7}\times S^{4}.$

\bigskip\ 

\noindent \textbf{4.- Englert solution revisited}

\smallskip\ 

In the case of Englert solution we have also $AdS_{4}\times S^{7},$ but (1)
or (6) are no longer satisfied and therefore the right hand side of the
first field equation in (23) is different from zero. This means that
according to the discussion of the previous section maximal supersymmetry is
broken. In turn, this implies that $F^{A_{1}A_{2}A_{3}A_{4}}$ cannot be
written in the form (22) or in other words $F^{A_{1}A_{2}A_{3}A_{4}}$ does
not satisfy (21). However we would like to emphasize that in spite of $%
F^{A_{1}A_{2}A_{3}A_{4}}$ does not satisfy (21) the background solution is
still the same as the Freund-Rubin solution, namely $AdS_{4}\times S^{7}$.
How is this possible? The answer comes from one of the division algebras:
the octonionic structure.

Consider the octonion identity

\begin{equation}
f^{abcd}f_{efgd}=\delta _{e}^{[a}\delta _{f}^{b}\delta _{g}^{c]}+\frac{1}{4}%
f_{[ef}^{[ab}\delta _{g]}^{c]}.  \tag{24}
\end{equation}%
with the indices $a,b,...etc$ running from $4$ to $11$. Here, $f_{abcd}$ is
a self dual object. Furthermore, $f_{abcd}$ is defined in terms of the
octonion structure constants $\psi _{ijk}$ and its dual $\varphi _{ijkl}$
through the relations

\begin{equation}
f_{ijk11}=\psi _{ijk}  \tag{25}
\end{equation}%
and

\begin{equation}
f_{ijkl}=\varphi _{ijkl}.  \tag{26}
\end{equation}%
From (24) it is not difficult to see that

\begin{equation}
f_{[ijk}^{r}f_{lmn]r}=0.  \tag{27}
\end{equation}%
This expression can be understood as a solution for

\begin{equation}
f_{s[ijk}f_{lmn]r}=0,  \tag{28}
\end{equation}%
which remains us the formula (15) reduced to seven dimensions. In fact,
following a G\"{u}rsey and Tze [6], introducing a sieben-bein $h_{k}^{i}$
one can make this identification

\begin{equation}
F_{ijkl}=h_{i}^{r}h_{j}^{s}h_{k}^{t}h_{l}^{m}f_{rstm}  \tag{29}
\end{equation}%
and therefore (28) leads to

\begin{equation}
F_{s[ijk}F_{lmn]r}=0.  \tag{30}
\end{equation}%
Starting from (24) and following similar arguments we may establish that

\begin{equation}
F_{s[ijk}F_{lmnr]}=0  \tag{31}
\end{equation}%
and

\begin{equation}
F_{[sijk}F_{lmnr]}=0.  \tag{32}
\end{equation}%
Thus, according to the discussion of previous sections (30)\ and (31) imply
that $F_{ijkl}$ satisfies the relation

\begin{equation}
F_{sij[k}F_{lmnr]}=0  \tag{33}
\end{equation}%
which means that $F_{ijkl}$ is decomposable.

On the other hand, in four dimensions as we already mentioned, we can take

\begin{equation}
F^{\mu \nu \alpha \beta }=\lambda \varepsilon ^{\mu \nu \alpha \beta }, 
\tag{34}
\end{equation}%
where $\lambda $ is an arbitrary function. Since $\varepsilon ^{\mu \nu
\alpha \beta }$ is a maximal completely antisymmetric object in four
dimensions we get the formula

\begin{equation}
F_{\mu \nu \alpha \lbrack \beta }F_{\sigma \rho \tau \gamma ]}=0,  \tag{35}
\end{equation}%
which implies

\begin{equation}
F_{[\mu \nu \alpha \beta }F_{\sigma \rho \tau \gamma ]}=0.  \tag{36}
\end{equation}%
Thus, $F^{\mu \nu \alpha \beta }$ is also decomposable.

Our main observation is that despite both $F_{ijkl}$ and $F_{\mu \nu \alpha
\beta }$ are decomposable the eleven-dimensional four-form $F_{ABCD}$ is
not. The reason comes from the fact that if $F_{ijkl}$ and $F_{\mu \nu
\alpha \beta }$ are decomposables and the only nonvanishing components of $%
F_{ABCD},$ the relation $F_{A_{1}A_{2}A_{3}[A_{4}}F_{B_{1}B_{2}B_{3}B_{4}]}$
is different from zero and therefore the full $F_{ABCD}$ is not
decomposable. The result follows from the expression

\begin{equation}
F_{\mu \nu \alpha \lbrack \beta }F_{ijkm]}\neq 0,  \tag{37}
\end{equation}%
or

\begin{equation}
F_{[\mu \nu \alpha \beta }F_{ijkm]}\neq 0.  \tag{38}
\end{equation}%
In fact, since $\varepsilon ^{\mu \nu \alpha \beta }$ and $f^{ijkm}$ take
values in the set $\{-1,0,1\}$ in general we have that

\begin{equation}
\varepsilon _{\mu \nu \alpha \lbrack \beta }f_{ijkm]}\neq 0,  \tag{39}
\end{equation}%
or

\begin{equation}
\varepsilon _{\lbrack \mu \nu \alpha \beta }f_{ijkm]}\neq 0.  \tag{40}
\end{equation}%
In turn this means that $F_{[A_{1}A_{2}A_{3}A_{4}}F_{B_{1}B_{2}B_{3}B_{4}]}%
\neq 0$ or $F\wedge F\neq 0$ and, according to the discussion of section 2,
consequently we no longer have maximal supersymmetric solution.
Nevertheless, as Englert showed, although the right hand side of the first
field equation in (23) is not vanishing the field equations still admit the
solution $AdS_{4}\times S^{7}.$

\bigskip\ 

\noindent \textbf{5.- Connection with chirotopes}

\smallskip\ 

The aim of this section is to discuss the formalism described in section 2,
3 and 4 from the point of view of the oriented matroid theory. Indeed, our
discussion will focus on the chirotope concept which provides one possible
definition of an oriented matroid.

Chirotopes had been a major subject of investigation in mathematics during
the last 25 years [1]. Roughly speaking a chirotope is a combinatorial
abstraction of subdeterminants of a given matrix. More formally, a
realizable $p$-rank chirotope is an alternating function $\chi
:\{1,...,n\}^{p}\rightarrow \{-1,0,1\}$ satisfying the Grassmann-Pl\"{u}cker
relation

\begin{equation}
\chi _{\hat{A}_{1}...\hat{A}_{n-1}[\hat{A}_{p}}\chi _{\hat{B}_{1}...\hat{B}%
_{p}]}=0,  \tag{41}
\end{equation}%
while nonrealizable $p$-rank chirotope corresponds to the case

\begin{equation}
\chi _{\hat{A}_{1}...\hat{A}_{n-1}[\hat{A}_{p}}\chi _{\hat{B}_{1}...\hat{B}%
_{p}]}\neq 0.  \tag{42}
\end{equation}%
It is worth mentioning that there is a close connection between chirotopes
and Grassmann variety. In fact, the Grassmann-Pl\"{u}cker relations describe
a projective embedding of the grassmannian of planes via decomposable $p$%
-forms (see Ref. [1] for details).

Thanks to our revisited review of Freund-Rubin and Englert solutions given
in the previous sections we find that the link between this these solutions
and the chirotope is straightforward. In fact, our first observation is that
any $\varepsilon $-symbol is in fact a realizable chirotope (see Refs. [12]
and [7]), since it is always true that

\begin{equation}
\varepsilon _{\hat{A}_{1}...\hat{A}_{n-1}[\hat{A}_{n}}\varepsilon _{\hat{B}%
_{1}...\hat{B}_{n}]}=0.  \tag{43}
\end{equation}%
From this perspective we recognize that the formula (21) indicates that in
the case of maximal supersymmetry the four-form $F_{ABCD}$ is a realizable $%
4 $-rank chirotope. While in the case of Englert solution, from (39) and
(42) we discover that one may identify $F_{ABCD}$ with a nonrealizable $4$%
-rank chirotope. We think that this identification open the possibility to
introduce other chirotopes no necessarily related to octonions as a solution
for eleven-dimensional gravity.

\bigskip\ 

\noindent \textbf{6.- Final Remarks}

\smallskip\ 

We have identified the Freund-Rubin-Englert solution for eleven-dimensional
supergravity with the chirotope concept. In the case of maximally
supersymmetric solution the four-form $F$ can be identified with a
realizable $4$-rank chirotope, while in the case of Englert solution, $F$
may correspond to a nonrealizable $4$-rank chirotope. However, there are
many possible $4$-rank chirotopes in eleven dimensions and therefore there
must be many new and unexpected solutions for eleven dimensional gravity.

One of our key tools in our formalism is the octonionic structure. This
division algebra was already related to the Fano matroid and therefore, a
possible connection with supergravity was established in Ref [17]. Here, we
have been more specific and through the chirotope concept we established the
relation between the Freund-Rubin-Englert solution and oriented matroid
theory. However, it may be interesting to understand the possible role of
the Fano matroid in this scenario.

Here, we focused on eleven-dimensional supergravity but according to the
results given in Ref. [6] in principle, one may expect to apply similar
procedure in the case of ten-dimensional supergravity and other higher
dimensional supergravities such as Type I supergravity and massive IIA
supergravity.

An important property in the oriented matroid theory is that one can
associate any chirotopes with its dual. Thus, working on the framework of
oriented matroids we can assure that any possible solution for
eleven-dimensional gravity in terms of chirotopes shall have a dual
solution. This means that this kind of solution contains automatically a
dual symmetry.

Using the idea of matroid bundle [18]-[22], Guha [23] has observed that
chirotopes can be related to Nambu-Poisson structure. It may be interesting
for further research to see whether the present formalism can be useful to
bring the Nambu-Poisson structure to eleven-dimensional supergravity.

\bigskip\ 

\noindent \textbf{Acknowledgment: }I would like to thank L. Ruiz and J.
Silvas for their helpful comments.

\smallskip\

\end{document}